\documentclass[aps,prd,nofootinbib,twocolumn,superscriptaddress,floatfix,notitlepage]{revtex4-1}
\pdfoutput=1
\usepackage{graphicx}
\usepackage{amsmath,amsfonts,amssymb}
\usepackage{color}
\usepackage[breaklinks,colorlinks,urlcolor=blue,citecolor=blue,linkcolor=magenta]{hyperref}
\usepackage{verbatim}
\usepackage{enumitem}
\usepackage{hyperref}
\usepackage{aas_macros}

\newcommand{\be}{\begin{equation}} 
\newcommand{\ee}{\end{equation}}
\newcommand{\bea}{\begin{equation}\begin{aligned}} 
\newcommand{\eea}{\end{aligned}\end{equation}}

\newcommand{\td}{{\rm d}}
\newcommand{\Mpc}{{\rm Mpc}}
\newcommand{\Msun}{M_\odot}

\def\lsim{\mathrel{\raise.3ex\hbox{$<$\kern-.75em\lower1ex\hbox{$\sim$}}}}
\def\gsim{\mathrel{\raise.3ex\hbox{$>$\kern-.75em\lower1ex\hbox{$\sim$}}}}

\begin{document}

\title{Did JWST observe imprints of axion miniclusters or primordial black holes?}

\author{Gert H\"utsi}
\email{gert.hutsi@kbfi.ee}
\affiliation{Keemilise ja Bioloogilise F\"u\"usika Instituut, R\"avala pst. 10, 10143 Tallinn, Estonia}
\author{Martti Raidal}
\email{martti.raidal@cern.ch}
\affiliation{Keemilise ja Bioloogilise F\"u\"usika Instituut, R\"avala pst. 10, 10143 Tallinn, Estonia}
\author{Juan Urrutia}
\email{juan.urrutia@kbfi.ee}
\affiliation{Keemilise ja Bioloogilise F\"u\"usika Instituut, R\"avala pst. 10, 10143 Tallinn, Estonia}
\author{Ville Vaskonen}
\email{ville.vaskonen@pd.infn.it}
\affiliation{Keemilise ja Bioloogilise F\"u\"usika Instituut, R\"avala pst. 10, 10143 Tallinn, Estonia}
\affiliation{Dipartimento di Fisica e Astronomia, Universit\`a degli Studi di Padova, Via Marzolo 8, 35131 Padova, Italy}
\affiliation{Istituto Nazionale di Fisica Nucleare, Sezione di Padova, Via Marzolo 8, 35131 Padova, Italy}
\author{Hardi Veerm\"ae}
\email{hardi.veermae@cern.ch}
\affiliation{Keemilise ja Bioloogilise F\"u\"usika Instituut, R\"avala pst. 10, 10143 Tallinn, Estonia}

\begin{abstract}
The James Webb Space Telescope has detected surprisingly luminous early galaxies that indicate a tension with the $\Lambda$CDM. Motivated by scenarios including axion miniclusters or primordial black holes, we consider power-law modifications of the matter power spectrum. We show that the tension could be resolved if dark matter consists of $2\times 10^{-18}{\rm eV}$ axions or if a fraction $f_{\rm PBH} > 0.005$ of dark matter is composed of compact heavy $4\times 10^6 M_\odot (f_{\rm PBH}/0.005)^{-1}$ structures such as primordial black hole clusters. 
However, in both cases, the star formation efficiency needs to be significantly enhanced.
\end{abstract}

\maketitle

\section{Introduction}

The long-awaited next-generation space telescope -- the James Webb Space Telescope (JWST)~\cite{JWST} -- is finally delivering data. Arguably the most intriguing result so far has been the detection of high-redshift galaxies with surprisingly high stellar masses~\cite{2022arXiv220802794N,2022arXiv220712446L,2023MNRAS.519.1201A,2022ApJ...940L..55F,2022arXiv220801612H,2022arXiv221001777B}. This is somewhat reminiscent of what happened almost three decades ago with the Hubble Space Telescope (HST)~\cite{HST} discovering the unexpected richness of galaxies in the Hubble Deep Field. These early hints indicated severe challenges for the then standard Einstein-de Sitter cosmology and together with subsequent observational data (most importantly cosmic microwave background (CMB), type Ia supernovae and large-scale structure data from the large redshift surveys) led to the establishment of the current standard cosmological model -- the $\Lambda$CDM model. If true, the new results from JWST might similarly call for a significant modification of the $\Lambda$CDM.

There have been several attempts in trying to make the concordance $\Lambda$CDM cosmology compatible with JWST measurements. For example, reduced dust attenuation in high-redshift galaxies makes them appear brighter~\cite{Ferrara:2022dqw,Ziparo:2022rir}. However, it seems that altering the dust production alone is not enough to relieve these tensions -- one also has to make the star formation significantly more effective~\cite{2023MNRAS.519..843M}. Even then, by pushing the star formation efficiency to the extremes, it is still hard to accommodate these new observations within the $\Lambda$CDM cosmology~\cite{Boylan-Kolchin:2022kae,Lovell:2022bhx}. Moreover, a high star formation efficiency leads to a high abundance of ionizing photons that may contradict the measurements of the cosmic reionization history. A potential solution was proposed in~\cite{Gong:2022qjx} within fuzzy DM models that suppress the abundance of small-scale structures.

Although there are significant uncertainties involved in pushing the Lyman-break technique to so high redshifts -- difficulties in interpretation of spectral breaks or additional contamination due to emission lines~\cite{2022arXiv220812828F,2022arXiv220814999E} -- in this paper we assume these observational inferences to be true and investigate beyond the $\Lambda$CDM physics that could resolve the tension. Different modifications to the small-scale physics boosting the abundance of $\mathcal{O}(10^{11}) \Msun$ dark matter (DM) halos at high redshifts $z > 9$ have been already proposed. These proposals include heavy primordial black holes (PBHs)~\cite{Liu:2022bvr}, non-Gaussianities in primordial fluctuations~\cite{Biagetti:2022ode}, and modified dark energy equation of state~\cite{Menci:2022wia}.

In this paper, we focus on scenarios, including axion miniclusters or heavy PBHs, which boost the matter power spectrum at $k > 3 h \Mpc^{-1}$. Our analysis uses extreme value statistics (EVS)~\cite{Gumbel1958,Coles2001} and considers two of the most extremal galaxies observed so far by JWST: CEERS-1749 of stellar mass $\log_{10}(M_*/\Msun) = 9.6\pm0.2$ at redshift $z = 16\pm 0.6$~\cite{2022arXiv220802794N} and Galaxy 14924 of stellar mass $\log_{10}(M_*/\Msun) = 10.9\pm0.3$ at redshift $z = 9.9\pm 0.5$~\cite{2022arXiv220712446L}.

\section{Matter power spectrum}

The matter power spectrum is strictly constrained at scales $k \lsim 3 h \Mpc^{-1}$ by measurements of galaxy clustering, Lyman-alpha forest data and the UV luminosity function~\cite{Reid:2009xm,Chabanier:2019eai,Sabti:2021unj}. The latter is obtained using data from the HST and will improve with the JWST observations~\cite{Sabti:2021xvh,2023MNRAS.518.6011D}. These observations agree well with the predictions of the standard $\Lambda$CDM model. At smaller scales, significant deviations from the CDM matter power spectrum are possible and affect the abundance of $M\lsim 10^{11} \Msun$ halos. In particular, a matter power spectrum that exceeds the $\Lambda$CDM prediction at scales $\mathcal{O}(10) h \Mpc^{-1}$ gives rise to an increased abundance of massive early galaxies, perhaps being compatible with the JWST observations.

\begin{figure}
\centering
\includegraphics[width=0.95\columnwidth]{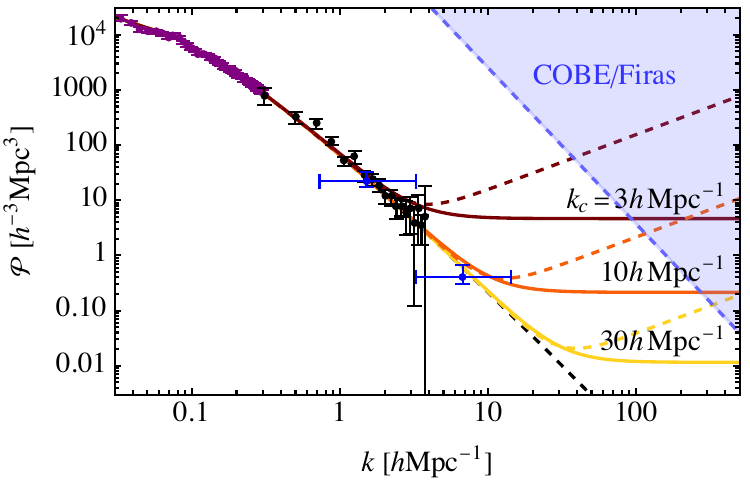}
\caption{The matter power spectrum for standard $\Lambda$CDM (black dashed) and for three values of $k_c$ with $n=0$ (solid) and $n=1$ (dashed). The points with errors correspond to SDSS measurements of galaxy clustering from luminous red galaxies (purple)~\cite{Reid:2009xm} and Lyman-$\alpha$ forest (black)~\cite{Chabanier:2019eai} and HST measurements of UV luminosity function (blue)~\cite{Sabti:2021unj}. The blue dashed curve shows the maximal cut-off scale for $k^4$ growth of the adiabatic curvature power spectrum from COBE/Firas bound on $\mu$ and $y$ distortions~\cite{Chluba:2013dna}.}
\label{fig:Pk}
\end{figure}

We consider power-law modifications to the matter power spectrum at small scales:
\be \label{eq:Pk}
    \mathcal{P}(k) = \mathcal{P}_{\rm CDM}(k) + \mathcal{P}_{\rm CDM}(k_c) \left(\frac{k}{k_c}\right)^n \,,
\ee
where $\mathcal{P}_{\rm CDM}(k)$ denotes the standard $\Lambda$CDM matter power spectrum, $k_c > 3 h \Mpc^{-1} $ the scale above which the power-law behaviour takes over and $n$ the spectral index of the power-law. The power-law part is cut at some scale $k_{\rm cut} > k_c$. As discussed later, our results are insensitive to the exact value as long as $k_{\rm cut} \gtrsim 30 h \Mpc^{-1}$. For the CDM matter power spectrum we consider a nearly scale-invariant curvature power spectrum with the spectral index $n_s = 0.96$ and the fitted transfer function from Ref.~\cite{Eisenstein:1997ik}. In Fig.~\ref{fig:Pk} we show the CDM matter power spectrum and examples of the modified matter power spectra~\eqref{eq:Pk} together with the constraints from measurements of galaxy clustering~\cite{Reid:2009xm}, Lyman-alpha forest data~\cite{Chabanier:2019eai} and UV luminosity function~\cite{Sabti:2021unj}.

Deviations of the form of Eq.~\eqref{eq:Pk} from the standard $\Lambda$CDM matter power spectrum can arise, for example, in the following scenarios:
\begin{enumerate}[leftmargin=*]
    \item In axion-like particle DM models, in which the 
    global $U(1)$ symmetry is broken after the cosmic inflation, large amplitude small-scale fluctuations are generated via the Kibble mechanism leading to the formation of axion miniclusters~\cite{Kolb:1993zz}. This corresponds to an $n=0$ contribution to the matter power spectrum with an amplitude~\cite{Fairbairn:2017sil,Feix:2019lpo,Irsic:2019iff,OHare:2021zrq,Ellis:2022grh}
    \be
        A = \frac45 \frac{6 \pi^2}{D(z_{\rm eq})^2}  k_{\rm cut}^{-3} \,,
    \ee
    where $D(z_{\rm eq})$ denotes the growth factor (normalized so that $D(0)=1$) at matter-radiation equality and $k_{\rm cut}$ the cut-off given by the scale that entered the horizon at the moment when the axion of mass $m_a$ began to oscillate (i.e., $3H = m_a$ at $aH = k_{\rm cut}$),\footnote{We assume, for simplicity, that the axion mass is temperature independent. For further discussion, see e.g. Refs.~\cite{Fairbairn:2017sil,OHare:2021zrq}.}
    \be
        k_{\rm cut} \simeq 300\, \Mpc^{-1} \sqrt{\frac{m_a}{10^{-18}\,{\rm eV}}} \,.
    \ee
    Using a power law approximation for $\mathcal{P}_{\rm CDM}$ around $k_c = \mathcal{O}(10) h\Mpc^{-1}$, we find that the scale $k_c$, at which $\mathcal{P}_{\rm CDM}(k_c) = A$, is given by 
    \be\label{eq:kc_AMC}
        k_c \approx 3 h \Mpc^{-1} \left(\frac{m_a}{10^{-18}\,{\rm eV}}\right)^{0.6} \,,
    \ee
    thus $k_{\rm cut} \gg k_c$ is satisfied in the relevant $k_c$ range.

    \item In models with heavy PBHs the graininess associated with the discrete Poisson distribution of PBHs~\cite{1975A&A....38....5M,Carr:2018rid} or PBH clusters~\cite{DeLuca:2022uvz} (but possibly also other compact structures) generates an $n=0$ contribution to the matter power spectrum with an amplitude~\cite{Inman:2019wvr,DeLuca:2020jug}
    \bea
        A 
        = \frac{1}{n_{\rm PBH}} \frac{f_{\rm PBH}^2}{D(z_{\rm eq})^2}
        = \frac{6\pi^2 f_{\rm PBH}^2}{D(z_{\rm eq})^2} k_{\rm cut}^{-3} \,,
    \eea
    where $f_{\rm PBH}$ denotes the fraction of DM in PBHs, $m_{\rm PBH}$ the PBH mass and $\rho_{\rm DM}$ the DM density. The cut-off scale of the spectrum is given by the average separation of PBHs,
    \bea
        k_{\rm cut} 
        &\simeq (6\pi^2 n_{\rm PBH})^{1/3} \\
        &= 900 h \Mpc^{-1} \left(\frac{f_{\rm PBH} \, 10^4 \Msun}{m_{\rm PBH}}\right)^{1/3}\,.
    \eea
    Below this scale, one expects a single PBH in the volume of the corresponding comoving sphere and thus the seed effect~\cite{1975A&A....38....5M,1983ApJ...275..405F,1983ApJ...268....1C,Carr:2018rid,Cappelluti:2021usg} begins to dominate over the Poisson one.\footnote{Importantly, the usual Press-Schechter formalism breaks down above $k_{\rm cut}$ as the fluctuations in the PBH density are non-Gaussian. Note that the scale at which the PBH-induced density fluctuations become non-linear, $k_{\rm cut, NL} \approx (n_{\rm PBH}/f_{\rm PBH})^{1/3} > k_{\rm cut}$ is smaller.} The scale $k_c$ is
    \be
        k_c \approx 6 h \Mpc^{-1} \left(\frac{f_{\rm PBH} m_{\rm PBH}}{10^4 \Msun}\right)^{-0.4}\, .
    \ee
    As Eq.~\eqref{eq:kc_AMC}, the latter approximation works well at $k_c = \mathcal{O}(10) h\Mpc^{-1}$. We find that $k_c < k_{\rm cut}$ if $f_{\rm PBH} > 10^{-4} (m_{\rm PBH}/10^4 \Msun)^{-0.09}$. Because CMB observations constrain $f_{\rm PBH} < 10^{-8}$ for $m_{\rm PBH} \gtrsim 10^4 \Msun$~\cite{Serpico:2020ehh}\footnote{For a review on constraints on PBHs, see e.g.~\cite{Carr:2020gox}.}, explanations for the JWST observations relying on the Poisson effect from heavy PBHs are not viable. However, the accretion constraint can be softened if lighter PBHs were formed in dense clusters (as described e.g. in~\cite{Khlopov:1998nm,Khlopov:2004sc,Flores:2020drq,Jung:2021mku,DeLuca:2021mlh}) with masses above $10^4 \Msun$. Such clusters would then act as sources for the Poisson fluctuations instead of the individual PBHs~\cite{DeLuca:2022uvz}. 
    
    We must stress that here we focused on the Poisson effect. As massive BHs may play an important role in the formation of the first galaxies~\cite{Silk:1997xw}, we cannot rule out the so-called seed effect from PBHs with $f_{\rm PBH} \leq 10^{-8}$ as a potential explanation. Assuming a monochromatic mass distribution, the expected number of PBHs of initial mass $M_{\rm PBH}$ in the JWST light cone (see Eq.~\eqref{eq:N_halo}) at redshifts $15<z<17$ is
    \be
        N_{{\rm PBH}, 15<z<17} = 2.8 \times \left(\frac{f_{\rm PBH}}{10^{-9}}\right) \left(\frac{M_{\rm PBH}}{10^{6} M_{\odot}}\right)^{-1}\, .
    \ee
    However, the number of massive PBHs cannot be straightforwardly translated into the number of luminous galaxies (such as CEERS-1749). Due to its non-linear nature and dependence on uncertain baryonic physics, the PBH seed scenario requires a dedicated study.
    
    \item Enhanced adiabatic power spectra can be generated, for example, in single field inflation if the inflaton potential has features which slow down the inflaton field~\cite{Garcia-Bellido:2017mdw,Kannike:2017bxn,Germani:2017bcs}. Such scenarios are relevant for PBH formation~\cite{Hawking:1971ei,Carr:1974nx} and are thus mostly studied in that context. However, they could be realized without any relation to PBHs. In typical models of single-field inflation, a curvature power spectrum with a constant spectral index $n_s$ at $k < k_c$ can grow as $k^{5-|2-n_s|}$ at $k>k_c$~\cite{Karam:2022nym}. For $n_s = 0.96$ this translates in Eq.~\eqref{eq:Pk} to $n = 0.68$ for $k_c\simeq 1h\,\Mpc^{-1}$ and $n = 0.14$ for $k_c\gg 100h\,\Mpc^{-1}$. Although an $k^4$ growth of the curvature power spectrum is typical~\cite{Byrnes:2018txb}, steeper spectra can be realized in inflationary scenarios with increasingly fine-tuned features~\cite{Tasinato:2020vdk}. In the following, we consider at most $n=1$.
\end{enumerate}

In the first two cases, the power-law contribution to the matter power spectrum constitutes an isocurvature component to the matter fluctuations. The COBE/Firas constraint shown in Fig.~\ref{fig:Pk} applies to the adiabatic component, while the constraint from $\mu$ and $y$ distortions on the isocurvature component is much weaker~\cite{Chluba:2013dna}. In particular, the cut-off scales in the first two scenarios are far below that constraint. 

In the third case, the fluctuations are adiabatic, and the COBE/Firas constraint excludes PBH formation if $k_c \lsim 10^2 h \Mpc^{-1}$ unless the growth of the curvature power spectrum is relatively slow, less than $\propto k^3$, which translates to $n<0$ in Eq.~\eqref{eq:Pk}. However, steep growth of the curvature power spectrum at scales $k_c = \mathcal{O}(10)h\Mpc^{-1}$ that terminates with amplitude less than four orders of magnitude above the CMB amplitude is not excluded.

\begin{figure}
\centering
\includegraphics[width=\columnwidth]{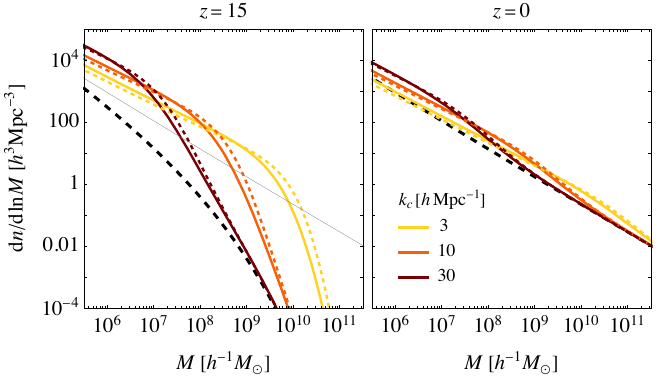}
\caption{The halo mass function for the matter power spectra shown in Fig.~\ref{fig:Pk} at redshifts $z=15$ and $z=0$. The thin line shows the CDM halo mass function at $z=0$.}
\label{fig:hmf}
\end{figure}

\section{Halo mass function}

The halo mass function can be expressed as
\be
    \frac{\td n}{\td \ln M} = \frac{\rho_m}{M} \nu f(\nu) \frac{\td \ln \nu}{\td \ln M} \,,
\ee
where $\rho_m$ denotes the average present mass density and $\nu \equiv \delta_c(z)^2/\sigma_M^2$. The critical overdensity required for collapse is $\delta_c(z) = 1.686/D(z)$, where $D(z)$ is the growth factor~\cite{Dodelson:2003ft},\footnote{This growth factor is for adiabatic fluctuations. While it is slightly different for isocurvature fluctuations~\cite{Inman:2019wvr}, we find that this does not make a significant difference in our results.}
\be
    D(z) \propto H(z) \int_z^\infty \!\td z'\, \frac{(1+z')}{H(z')^3} \,,
\ee
normalized such that $D(0)=1$. In the Press-Schechter formalism~\cite{Press:1973iz} with corrections from ellipsoidal dynamics~\cite{Sheth:1999mn}
\be
    \nu f(\nu) = A \left[ 1+(q\nu)^{-p} \right] \sqrt{\frac{q\nu}{2\pi}} e^{-q\nu/2} \,,
\ee
where $p = 0.30$, $A = 0.322$ and $q = 0.75$~\cite{Cooray:2002dia}. The variance $\sigma_M^2$ at the mass scale $M$ is 
\be
    \sigma_M^2 = \frac{1}{2\pi^2} \int \td k\, k^2 \mathcal{P}(k) W^2(kR) \big|_{R = R(M)} \,,
\ee
and the comoving smoothing scale $R$ is related to the halo mass by $M = 4\pi \rho_m R^3/3$. We use a real space top-hat window function $W(x) = 3(\sin x - x \cos x)/x^3$.

In Fig.~\ref{fig:hmf} we show the halo mass function corresponding to the modified power spectra shown in Fig.~\ref{fig:Pk}. We see that the deviation of the halo mass function from the standard $\Lambda$CDM prediction at the relevant scales depends mostly on the scale $k_c$ and less on the slope $n$. This indicates that the exact shape of the modification in $\mathcal{P}$ has a relatively minor impact on our results and that the two-parameter model~\eqref{eq:Pk} is sufficient to capture the relevant qualitative features. Moreover, we find that the halo mass function at the relevant mass scales does not significantly depend on the cut-off scale of the power-law part as long as $k_{\rm cut} \gtrsim 30 h \Mpc^{-1}$.

\begin{figure}
\centering
\includegraphics[width=0.98\columnwidth]{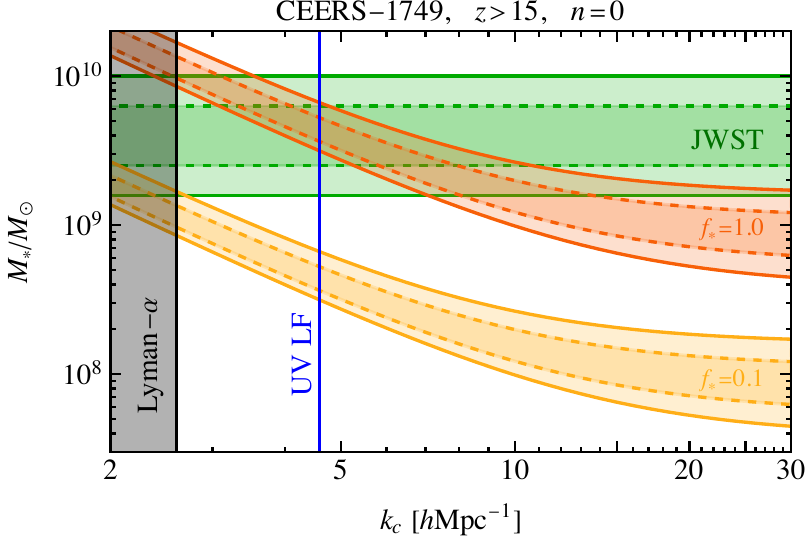}
\caption{The yellow and red bands show the $1\sigma$ (dashed) and $2\sigma$ (solid) stellar mass EVS confidence intervals at $z>15$ for star formation efficiencies $f_*=0.1$ (yellow) and $f_*=1.0$ (red) and the spectral index $n=0$. The gray region is excluded at $2\sigma$ by the constraints on the matter power spectrum from Lyman-$\alpha$ clustering~\cite{Chabanier:2019eai} and the blue line shows the conservative $2\sigma$ constraint from the UV luminosity function~\cite{Sabti:2021unj}. The green band indicates the stellar mass estimate of CEERS-1749 at $z > 15$ from~\cite{2022arXiv220802794N}.}
\label{fig:kM}
\end{figure}

\begin{figure*}
\centering
\includegraphics[width=0.94\linewidth]{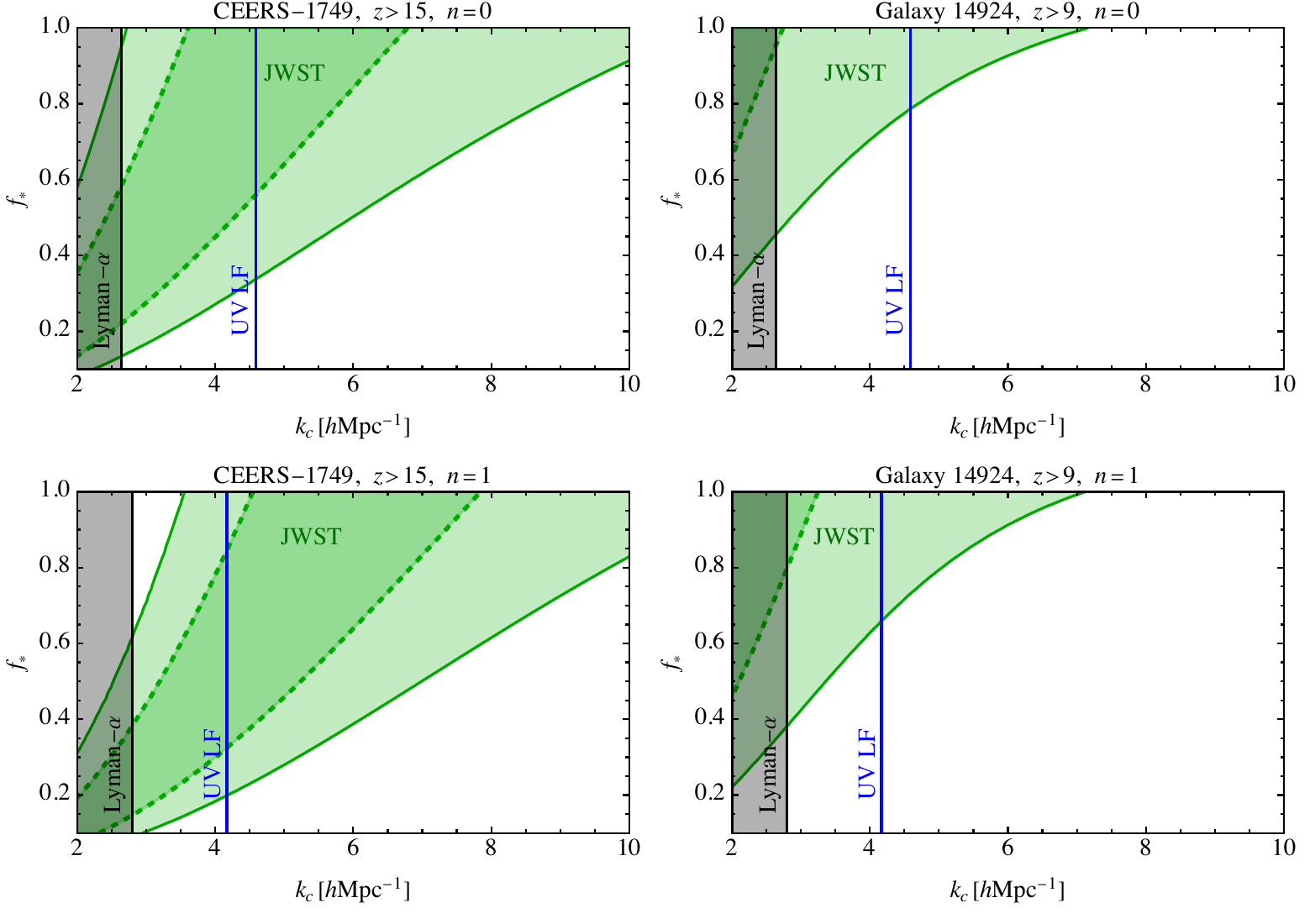}
\vspace{-4mm}
\caption{The green bands indicate the $1\sigma$ and $2\sigma$ confidence intervals of the likelihood averaged over the EVS PDF. The left panels are for the stellar mass estimate of CEERS-1749 at $z > 15$ and the right panels are for the stellar mass estimate of Galaxy 14924 at $z > 9$.}
\label{fig:jwst}
\end{figure*}

\section{Analysis and results}

The expected number of halos in the mass range $M_{\rm min}<M<M_{\rm max}$ observed in the redshift range $z_{\rm min} < z < z_{\rm max}$ is given by the light-cone integral
\be\label{eq:N_halo}
    N
    = f_{\rm sky} \int_{z_{\rm min}}^{z_{\rm max}} \!\td z \frac{\td V_c}{\td z} \int_{M_{\rm min}}^{M_{\rm max}} \!\!\td M \frac{\td n}{\td M} \,,
\ee
where $f_{\rm sky}$ is the fraction of the sky being observed and $V_c$ is the comoving Hubble volume at redshift $z$. For the JWST observations, the relevant quantity is the stellar mass of the halo, related to the halo mass $M$ via~\cite{Boylan-Kolchin:2022kae}
\be
    M_* = f_* f_b M \,,
\ee 
where $0<f_*<1$ is the star formation efficiency and $f_b\approx \Omega_b/\Omega_m\approx 0.16$ is the fraction of all matter in baryons. For example, taking the CDM halo mass function and a typical optimistic efficiency according to lower redshift observations $f_*=0.1$~\cite{2013ApJ...770...57B}, we find that the expected number of halos similar to CEERS-1749 whose stellar mass is $M_* > 10^{9.6} \Msun$ at redshift $z > 15$ within $f_{\rm sky} = 40\,{\rm arcmin}^2$ 
is $N \approx 7\times 10^{-8}$.

To compute the distribution of the heaviest halo expected to be seen in an observed light-cone volume we utilize EVS~\cite{2011MNRAS.418L..20H,2012MNRAS.421L..19H}. The EVS probability density function (PDF) in the full mass range $0<M<\infty$  is~\cite{2011MNRAS.418L..20H}
\be \label{eq:Phi}
    \Phi(M) = N f(M) F(M)^{N-1} \,,
\ee
where
\be
f(M) \propto \int_{z_{\rm min}}^{z_{\rm max}} \!\td z \frac{\td V_c}{\td z} \frac{\td n}{\td M} \,, \quad F(M) = \int_0^M \!\td M' f(M')
\ee
are the normalized PDFs of halo masses within the redshift range $z_{\rm min} < z < z_{\rm max}$ and the corresponding cumulative distribution.

In Fig.~\ref{fig:kM}, we show the $1\sigma$ and $2\sigma$ confidence intervals of the EVS PDF as a function of $k_c$ for $f_{\rm sky} = 2.7\times 10^{-7}$ and $z>15$. As expected, the distribution prefers larger stellar masses at smaller $k_c$. Moreover, the mean of the EVS distribution deviates by less than $10\%$ from the $M_*$ value corresponding to $N(M>M_*/f_*f_b, z>15)=1$ within $f_{\rm sky} = 2.7\times 10^{-7}$.~\footnote{Even though the widely used condition $N=1$ provides a reasonably accurate mass scale it does not provide similarly evident treatment for the accompanying stochasticity. This is the main reason to use EVS, which equips us with a complete statistical framework.} The green shaded region indicates the stellar mass of CEERS-1749, the gray shaded region is excluded by the Lyman-$\alpha$ constraints on the matter power spectrum and the blue line indicates the conservative constraint from the UV luminosity function.\footnote{The dominant UV luminosity function constraint $a_s = 0.66^{+0.43}_{-0.17}$ in Ref.~\cite{Sabti:2021unj} is given for the relative enhancement over the $\Lambda$CDM in the bin $0.5\,\Mpc^{-1} \leq k < 2.25\,\Mpc^{-1}$. As the constraint depends non-trivially on the shape of the power spectrum and has not been computed for the parametrization~\eqref{eq:Pk}, we estimate the corresponding constraint on $k_c$ by comparing the amplitude relative to the $\Lambda$CDM at the lower edge of the bin, $k=0.5\, \Mpc^{-1}$.} We see that with the commonly used efficiency $f_*=0.1$ the CEERS-1749 observation is in more than $2\sigma$ tension with the model even at the smallest allowed value of $k_c$. With the maximal efficiency $f_*=1.0$ a good agreement is evident if $k_c \lsim 10h{\rm kpc}^{-1}$.

The final results of our analysis are summarized in Fig.~\ref{fig:jwst}. The green bands show the $1\sigma$ and $2\sigma$ ranges in $k_c$ computed from the likelihood
\be
    \int \td M \, \Phi(M) p(M|M_c) \,,
\ee
assuming a log-normal distribution, $p(M|M_c) = \exp[-\ln(M/M_c)/2\sigma^2]/\sqrt{2\pi \sigma^2}$, for the measured mass $M_c$. The left panels show these ranges for CEERS-1749 and the right panels for Galaxy 14924. We find that a significantly enhanced star formation efficiency and a modification over the CDM matter power spectrum at scales $k_c = \mathcal{O}(10)h\Mpc^{-1}$ would be required to alleviate the tension below the $2\sigma$ level. A high star formation efficiency may lead to tensions with the CMB measurements of reionization history~\cite{Gong:2022qjx}. The lowest $f_*$, still compatible with both the Lyman-$\alpha$ and UV luminosity function constraints at the $2\sigma$ confidence level, is obtained for $k_c = 5h\Mpc^{-1}$. This can be reached if
\begin{enumerate}[leftmargin=*]
    \item DM consists of axions of mass $m_a\approx 2\times 10^{-18}\,{\rm eV}$ and the Peccei-Quinn symmetry is broken after inflation so that axion miniclusters are formed. The mass range $10^{-19}\,{\rm eV} \lsim m_a \lsim 10^{-16}\,{\rm eV}$ is, however, potentially excluded by black hole superradiance\footnote{The constraints can be alleviated in self-interacting scenario where the superradiant growth of the axionic cloud may be suppressed~\cite{Baryakhtar:2020gao}.} and the observed spins of superheavy black holes~\cite{Stott:2018opm,Unal:2020jiy,2168507}. For $m_a > 10^{-16}\,{\rm eV}$ we get $k_c > 50 h{\rm Mpc}^{-1}$, which does not affect the halo mass function at the relevant scales.
    
    \item a fraction $f_{\rm PBH}>0.005$ of DM consists of PBHs or PBH clusters of mass $m_{\rm PBH} \approx 4\times 10^6 \Msun (f_{\rm PBH}/0.005)^{-1}$. However, accretion constraints~\cite{Serpico:2020ehh} exclude the explanation from the Poisson effect from uniformly distributed individual PBHs, but the explanation with heavy clusters of subsolar mass PBHs may be viable.
    
    \item the curvature power spectrum grows steeply at $4h\Mpc^{-1}$ and the growth terminates before the COBE/Firas bound. This case is possible though not well motivated. In particular, it is non-trivial to realize such growth at those scales while simultaneously remaining in agreement with the CMB constraints.
\end{enumerate}

\section{Conclusions}

We have performed a careful analysis of whether a modified matter power spectrum could explain the existence of two of the most extreme galaxies observed so far by JWST, CEERS-1749 and Galaxy 14924. Considering power-law modifications to the matter power spectrum and extreme value statistics, we have estimated the distribution of the heaviest galaxies expected to be seen by JWST above a given redshift. We have found that both a high star formation efficiency and a modification of the matter power spectrum at the smallest scales allowed by the Lyman-$\alpha$ and UV luminosity function constraints are needed for this distribution to be compatible with the observed stellar mass, in particular for Galaxy 14924.

We have shown that extreme stellar masses consistent with the JWST observations can be obtained in cosmologies containing axion miniclusters, but the relevant range of axion masses is potentially excluded by superradiance constraints. Explanations relying on isocurvature perturbations from heavy Poisson-distributed PBHs are instead in conflict with existing accretion constraints on PBH abundance. However, such conflicts can be resolved in PBH scenarios with initial spatial clustering.

\vspace{10pt}
\begin{acknowledgments}
\vspace{-9pt}
We thank Diego Blas, Julian Mu{\~n}oz and Nashwan Sabti for useful discussions on the UV luminosity function. This work was supported by European Regional Development Fund through the CoE program grant TK133 and by the Estonian Research Council grant PRG803. The work of VV has been partially supported by the European Union's Horizon 2021 research and innovation program under the Marie Sk\l{}odowska-Curie grant agreement No. 101065736.
\end{acknowledgments}

\bibliography{refs}

\end{document}